\documentclass[aps,prd,a4paper,onecolumn,amsmath,showpacs,superscriptaddress,nofootinbib,preprintnumbers,notitlepage]{revtex4-1}
\usepackage{amssymb,amsmath}
\usepackage{graphicx}
\usepackage{color}
\usepackage{dcolumn}
\usepackage{pgfplots}
\usepackage{booktabs}
\usepackage{multirow}
\usepackage{dcolumn}
\usepackage{amsmath}
\usepackage{mathtools}
\usepackage{amsfonts}
\usepackage{amssymb}
\usepackage{epstopdf}
\usepackage{bm}
\usepackage{siunitx}
\usepackage{braket}
\usepackage{enumitem}
\usepackage{soul}
\usepackage{ulem}
\usepackage{color}
\usepackage{transparent}
\usepackage{pifont}
\usepackage[font={small}]{caption}

\begin{document}
\title{
Reconstructing inflation and reheating in  $f(\phi)T$ gravity }
\author{Ramón Herrera}
\email{ramon.herrera@pucv.cl}
\affiliation{Instituto de F\'{\i}sica, Pontificia Universidad Cat\'{o}lica de Valpara\'{\i}so, Avenida Brasil 2950, Casilla 4059, Valpara\'{\i}so, Chile.
}

\author{Carlos  Ríos}
\email{carlos.rios@ucn.cl}
\affiliation{Instituto de F\'{\i}sica, Pontificia Universidad Cat\'{o}lica de Valpara\'{\i}so, Avenida Brasil 2950, Casilla 4059, Valpara\'{\i}so, Chile.
}
\affiliation{Departamento de Ense\~nanza de las Ciencias B\'asicas, Universidad Cat\'olica del Norte, Larrondo 1281, Coquimbo, Chile.}

\begin{abstract}
In this work we study the reconstruction of an inflationary universe in the context of a theory of gravity $f(\phi)T$, in which  $T$ corresponds to the trace of energy momentum tensor. To realize this reconstruction during the inflationary epoch, we consider  as attractor the scalar spectral index $n_s$ in terms of  the number of $e$-folds $N$, in the framework of the slow-roll approximation. By assuming a specific function $f(\phi)$ together with the simplest attractor $n_s(N)$, we find different expressions  for the reconstructed  effective potential $V(\phi)$. Additionally, we analyze the era of reheating occurs  after of  the reconstruction obtained during the inflationary epoch. In this scenario we determine the duration and temperature during the reheating epoch, in terms of the equation of state parameter and of the observational parameters. In this context, the different parameters associated to the  reconstructed model are restricted during the scenarios of inflation and reheating by considering the recent astronomical observations. 

\end{abstract}

\maketitle
\section{Introduction}

It is well known that during the  evolution of the early universe, it exhibited  a short period of rapid growth called inflationary epoch or simply inflation \cite{1a,2a,2b}. In this context, the inflationary epoch provides  solutions to long standing cosmological problems associate to the  hot big bang model. However, the inflationary stage not only resolves the problematic  of the standard hot model, but also explicates  the large-scale structure (LSS) \cite{5a,6a},  as well  the anisotropies observed in the cosmic microwave background (CMB) radiation of the early universe \cite{7a,8a}.

In concern  to the different inflationary models that produce an 
adequate   evolution during the early universe, we can stand out those models that utilize   modifications to the Einstein's theory from a generalization in the  Lagrangian density of  Einstein-Hilbert action. In this sense, we can mention the function $f(R)$ in place of Ricci scalar $R$ in the Einstein-Hilbert action corresponds to a simple generalization of this action \cite{Buchdahl:1970ynr}, see also Refs.\cite{RR1,RR2}.   A modification to  the function $f(R)$ in the action  was developed in Ref.\cite{Bertolami:2007gv}, where the authors replace the function $f(R)$ by $f(R, \mathcal{L}_m)$,  in order   to include a coupling between an arbitrary function of the Ricci scalar and 
  the matter of the universe characterized by the matter Lagrangian density $\mathcal{L}_m$. In this context, different analysis in astrophysical and cosmology in relation to the non minimal coupling matter geometry coupling $f(R, \mathcal{L}_m)$ were developed in Refs.\cite{L1,L2}, see Ref.\cite{L3} for another formulations.
  
  In fact, we have another extensions  to the framework of standard General Relativity (GR) and in particular we distinguish the $f(R,\mathcal{L}_m,T)$ modified gravity or commonly called $f(R,T)$ gravity   in which the quantity $T$ denotes the trace of the energy momentum tensor  $T_{\mu\nu}$ related to the matter i.e., the trace  $T=g^{\mu\nu}\,T_{\mu\nu}$\cite{Harko:2011kv}, see also \cite{Houndjo:2011tu}. In this framework, an arbitrary function of the scalar Ricci $R$ as well as of the trace of the tensor $T_{\mu\nu}$ are considered to describe the early and present universe. The motivation to introduce the trace of the energy momentum tensor through the $f(R,T)$ gravity  comes from to consider some exotic matters or quantum effects
  to describe the universe. In the literature, different forms of the function $f(R,T)$ have been analyzed and in particular this function has been 
   decomposed  as combinations  of  arbitrary functions  related to  the Ricci scalar $\mathcal{R}(R)$  and of the trace of the energy momentum tensor $\mathcal{T}(T)$ such that;  $f(R,T)=\mathcal{R}(R)+\mathcal{T}(T)$ as well the multiplication  $f(R,T)=\mathcal{R}(R)\,\mathcal{T}(T)$. Another proposed theory along  the lines  $f(R,T)$ is  the energy momentum squared gravity proposed in Ref.\cite{Roshan:2016mbt}, in which the authors add to the standard Einstein-Hilbert action a quadratic term described by $T^2=T_{\mu\nu}\,T^{\mu\nu}$ in which as before   $T$ represents to the trace of the energy momentum tensor. In this context another modified gravity  analyzed in the literature associated to trace $T$ corresponds to add the $f(\phi)T$ term to the Einstein Hilbert action\cite{1}, where in this situation $f(\phi)$ is an arbitrary function of the scalar field (inflaton) coupled with the trace of the energy momentum tensor, see also recently Ref.\cite{Ashmita:2022dnv}. Here the authors studied the inflationary epoch (observational parameters) assuming  different effective potentials such as; chaotic inflation, natural inflation and the effective potential of the Starobinsky inflation obtained in the conformal frame. 
  In this context, the idea of  considering  an  extension of the modified gravity from the term $f(\phi,T)$ where the scalar field $\phi$ couples to the trace of energy momentum $T$   becomes interesting,  but also this extension offers an alternative approach to resurrect some inflationary models that do not work in the framework of the standard GR from the observational data such as; the chaotic model, natural inflation or other that are strongly disfavored from observations.  In this sense, the introduction of this extension of the  modified gravity transforms  the expressions associated to the observational parameters which are sensitive to $f(\phi,T)$ gravity.  
   
   Additionally, another models in which the trace $T$ plays an important  role during inflation, correspond to the models  related with $f(Q,T)$ gravity, in with $Q$ denotes the non-metricity scalar Refs.\cite{Xu:2019sbp,Xu:2020yeg} and for another extensions in which is considered the trace see e.g.,  Refs.\cite{Iosifidis:2021kqo,Harko:2021tav}.

On the other hand, the concept of  reconstruction associated to the physical variables that make up the background dynamics of the different  models during the inflationary scenario, considering the parameterization of   observational quantities such as; the scalar spectrum, scalar spectral index and the tensor to scalar ratio, have been analyzed by different authors \cite{P1,P2,P3,Chiba:2015zpa}. In particular, an interesting  reconstruction mechanism to find the physical quantities during inflation under the slow roll stage, corresponds to the parameterization of the scalar spectral index $n_s(N)$ and the tensor to scalar ratio $r(N)$ (called attractors) in terms of the  number of $e$-folds  $N$.  It is well known that the   parameterization for the scalar spectral index $n_s(N)$ as a function of the number of $e$-folds defined as  $n_s(N)=1-2/N$   is well supported  from observations  for values of  number $N\simeq$  50-70 from  the data taken by the Planck\cite{BICEP:2021xfz}.

In the theoretical  context  of the general relativity (GR), various inflationary models can be reconstructed under a single  parameterization or attractor given by $n_s\simeq 1-2/N$ assuming large $N$ 
such as; the hyperbolic tangent model or  T-model \cite{M1}, E-model\cite{M2}, $R^2$ or Starobinsky model\cite{1a} and  the famous model of  chaotic inflation\cite{2a}. However, the methodology used  for the reconstruction in the models  of warm  and Galileon inflation 
 was required  to introduce  two  attractors $n_s(N)$ and $r(N)$, in order to build the background variables \cite{Herrera:2018cgi,Herrera:2018mvo}, see also Ref.\cite{Gonzalez-Espinoza:2021qnv}. In the same way, it is possible to utilize the slow roll parameters $\epsilon(N)$ and $\eta(N)$ as a function of the number of $e$-folds $N$ to build the background variables and the observational parameters such as; the scalar spectrum index, power spectrum, tensor to scalar ratio among other\cite{Es1,Es2,Es3}. 
 For example, in Ref.\cite{Es1} was used some types of  parameterization for the slow roll  
 parameter $\epsilon(N)$ in order to find the effective potential in terms of the inflaton field. In the same way, in Ref.\cite{Es4} different effective potentials  were reconstructed by considering as parameterization the slow roll parameters   as a function of the number of $e$-folds $N$, see also \cite{Odintsov:2017fnc,Odintsov:2018ggm}.
 
In relation to the reheating of the universe, it is  known that  to recover the standard big-bang model, the early universe has to be reheated after of inflationary epoch. During the process of reheating
of the early universe, the components of  matter and radiation  are generated generally through the
decay of the scalar field or another fields, while the temperature of the universe increases
in magnitude and then the universe connects with the radiation epoch and then with 
the standard big-bang model\cite{St1}. However, there are various 
  reheating models (mechanisms) in order to increase the temperature during the early universe.  Thus, we have the mechanism of reheating in which from 
 the perturbative decay of an oscillating inflaton field at the end of inflationary epoch produces the reheating of the universe\cite{R1}, the mechanism associated to  
  non-perturbative processes as parametric resonance decay\cite{R2}, the reheating from tachyonic instability\cite{R3}, instant preheating in which this mechanism takes place   from  a non-perturbative processes and it occurs  almost instantly\cite{R4}, and also for  non oscillating models or called the NO models, in which the mechanism of reheating occurs from the another field  ``curvaton'' field (decay) \cite{R5}, see also Ref.\cite{yo2}.
 
 During  the reheating era we have  the different  parameters
 associated to the reheating. Thus,  we have that  this period 
 can be characterized by the reheating temperature $T_\text{reh}$, an effective equation of state (EoS)  $w_\text{reh}$ associated to the matter content in this process and one parameter related with the reheating duration and characterized   by  number of $e$-folds $N_\text{reh}$. 
 In relation to the reheating temperature a lower limit is restricted   by primordial nucleosynthesis (BBN) in which the temperature during the primordial nucleosynthesis $T_\text{BBN}\sim 10$ MeV, see e.g., \cite{El1}. In the context of the EoS parameter $w_\text{reh}$, we can mention  that different  numerical analysis were developed in order to characterize an effective EoS parameter from  specific interactions between the inflaton field and another matter fields, see e.g.,\cite{Podolsky:2005bw,Felder:2000hq}. Thus, we can consider that the EoS parameter $w_\text{reh}$ is a function of the cosmological time during the different scenarios of the reheating epoch. In this context, for example  for the canonical reheating stage assuming a  chaotic potential the EoS parameter at the end of inflation takes the value $w_\text{reh}=0$, but from the numerical analysis  the authors in Ref.\cite{Felder:2000hq} showed that this parameter increases to values of $w_\text{reh}\sim 0.3$ for cosmological time $t>200/M_\text{pl}$. In another scenarios such as a massive field the EoS parameter $w_\text{reh}$ increases from a negative value at the end of inflation 
 $w_\text{reh}=-1/3$\cite{Dodelson:2003vq} to $w_\text{reh}=0$, see e.g., \cite{Martin:2010kz,Munoz:2014eqa,Cook:2015vqa}. In this context, in a first approximation  we can assume that the EoS parameter $w_\text{reh}$ during the stage reheating can be considered as a constant in time through all the reheating epoch \cite{Munoz:2014eqa}.

The goal of this paper is to rebuild an inflationary model during the early universe  from a modified $f(\phi)T$  gravity. In this sense, we shall  analyze  an interaction between the scalar field $\phi$ and the trace of energy-momentum tensor $T$ of the form $f(\phi)T$, in order to reconstruct the inflationary stage  assuming the parameterization of the scalar spectral index $n_s$  as a function of the number of $e$-folds $N$ i.e., $n_s=n_s(N)$. In this framework, we study how the background dynamics in which there is an interaction between the field $\phi$ and the trace $T$ modifies  the reconstruction of the effective potential in terms of the scalar field
assuming as  attractor for large $N$ the scalar spectral index $n_s(N)$. Thus, from a general procedure, we will reconstruct the effective potential  from an attractor associated to the index $n_s(N)$ (for large-$N$). 

To reconstruct analytically  the effective potential in terms of the scalar field, we will analyze a specific example for the scalar spectral index parameterized in terms of $N$. In this form, we will assume the simplest attractor given by $n_s(N)=1-2/N$ for large-$N$. In this framework, we will rebuild the effective potential $V(\phi)$ and we will also obtain  the different constraints on the  parameters  associated to the reconstruction  (integration constants).

Additionally, we will study the reheating epoch from the reconstruction of the background variables obtained during the inflationary stage. In this context, we will determine the reheating parameters such as; the duration of the reheating from the number of $e$- folds, the temperature and the EoS during the reheating of the universe and how using the cosmological parameters  from Planck data (1$\sigma$ bound on $n_s$) these reheating parameters are constrained.

The outline of the article is as follows: the Sect. II we give
a brief description  of the modify gravity from an interaction between a coupling function that depends of the scalar field $f(\phi)$ and the trace of energy-momentum tensor $T$.
Here we analyze the background equations under the
slow-roll approximation and then we review the cosmological perturbations in this modify gravity. In Sect. III
we obtain, under a general formalism, explicit relations for the effective
scalar potential in terms of  the number of $e-$ folds $N$ to consider the reconstruction from the scalar spectral index $ns(N)$.
In the Sect. IV, we apply the reconstruction methodology in order to  find the scalar potential  V(N)  analytically. Besides,  we assume a specific case
in which we consider the simplest attractor for the spectral index
$n_s(N)=1-2/N$ for large $N$
together with a determined coupling function $f(\phi)$, in order to rebuild  the effective potential as a function of the scalar field. In Sect. V, we study the reheating scenario  for our model using the reconstructed potential obtained from the simplest attractor $n_s(N)$. Here, we determine the reheating temperature and the number of $e$-folds during the reheating era. 
Finally in Sect. VI we give our conclusions. We chose units so that
$c=\hbar=1$.

\section{The $f(\phi)T$ modified  gravity and the inflationary phase}

In this section, we shall consider the modified  gravity
from an interaction between the scalar field and the trace of the energy-momentum tensor through the term $f(\phi)T$. In this context,  
we start by writing down the action for this modified gravity as\cite{1}
\begin{equation}
\label{accion}
	S=\int\left[\frac{R}{2\kappa}+f(\phi)\,T+\mathcal{L}_m \right]\sqrt{-g}\,d^4\,x,
\end{equation}
where $R$ denotes the scalar Ricci, $g$ is the determinant of the metric $g_{\mu\nu}$ and  the constant $\kappa=8\pi\,G=M_{pl}^{-2}$ with $M_{pl}$ the Planck mass. The positive quantity  $f(\phi)$ is  an arbitrary function (dimensionless) associated to scalar field or inflaton field $\phi$, the expression  $\mathcal{L}_m$ denotes the matter Lagrangian density and   $T=g^{\mu\nu}T_{\mu\nu}$ corresponds to the trace of energy momentum tensor $T_{\mu\nu}$ associated to $\mathcal{L}_m$. In particular for the specific case in which the function  $f(\phi) \to 0$, the action given by Eq.(\ref{accion}) reduces to standard General Relativity (GR). 

It is well known that from the variation of the action (\ref{accion}) with respect to the metric, it gives rise to   the Einstein's equation;  
$
G_{\mu\nu}=R_{\mu\nu}-(1/2)g_{\mu\nu}R=\kappa\,\tilde{T}_{\mu\nu}
$, where the tensor $G_{\mu\nu}$ denotes the Einstein tensor and we have defined the tensor energy momentum $\tilde{T}_{\mu\nu}$ associated to the matter Lagrangian together with the coupling between the trace $T$ and the arbitrary function $f(\phi)$  such that
\begin{equation}
\label{TEMtilde}
	\tilde{T}_{\mu\nu}=\frac{-2}{\sqrt{-g}}\frac{\partial(\sqrt{-g}\tilde{\mathcal{L}}_m)}{\partial g^{\mu\nu}},
\end{equation} 
where  we have defined  $\tilde{\mathcal{L}}_m$ as  a combination of the matter Lagrangian density and the additional term   $ f(\phi) T$ such that   
\begin{equation}
	\tilde{\mathcal{L}}_m=f(\phi)\,T+\mathcal{L}_m.
\end{equation}

 In this form, we can write the tensor $\tilde{T}_{\mu\nu}$ given by Eq.(\ref{TEMtilde}) as 
\begin{equation}
	\tilde{T}_{\mu\nu}=	T_{\mu\nu}-2f(\phi)\left(T_{\mu\nu}-\frac{1}{2}g_{\mu\nu}T+\Pi_{\mu\nu}\right), \,\,\mbox{where}\,\,\,\,T_{\mu\nu}=\frac{-2}{\sqrt{-g}}\frac{\partial(\sqrt{-g}\mathcal{L}_m)}{\partial g^{\mu\nu}},	
\end{equation}
 corresponds to the   energy momentum  tensor related to the matter Lagrangian ${\mathcal{L}}_m$ and the quantity $\Pi_{\mu\nu}$ is defined as 
\begin{equation}
	\Pi_{\mu\nu}=g^{\alpha\beta}\frac{\partial T_{\alpha \beta}}{\partial g^{\mu\nu}}
	=-2T_{\mu\nu}+g_{\mu\nu}\mathcal{L}_m-2g^{\alpha\beta}\frac{\delta^2 \mathcal{L}_m}{\delta g^{\mu\nu}\delta g^{\alpha \beta}}\,\,.
\end{equation}
 
In the following we will assume  a single scalar field $\phi$ for the matter, in order to analyze the reconstruction of our inflationary model in the framework of $f(\phi)T$ gravity. In this sense,  we consider  that the  matter Lagrangian density $\mathcal{L}_m$ can be written as
\begin{equation}
	\mathcal{L}_m=-\frac{1}{2}g^{\mu\nu}\partial_\mu\phi\partial_\nu\phi-V(\phi),
\end{equation}
where $V(\phi)$ corresponds to the effective potential associated to scalar field. By assuming an inflaton field homogeneous i.e., $\phi=\phi(t)$ and considering a  perfect fluid of the form $\tilde{T}_{\mu}^{\nu}=$diag$(-\tilde{\rho},\tilde{p},\tilde{p},\tilde{p})$, in which $\tilde{\rho}$ and $\tilde{p}$ denote the effective  energy density and pressure, then we can identify these densities as
\begin{equation}
	\tilde{\rho}=\frac{1}{2}\dot{\phi}^2\left[1+2f(\phi)\right]+\left[1+4f(\phi)\right]V,
\end{equation}	
and 	
\begin{equation}	
	\tilde{p}=\frac{1}{2}\dot{\phi}^2\left[1+2 f(\phi)\right]-\left[1+4f(\phi) \right]V,
\end{equation}
where the dots mean derivatives with respect to the cosmological time and we have assumed the metric signature $(-,+,+,+)$.

In order to find the equation of motion of the scalar field we can utilize the continuity equation 
$\dot{\tilde{\rho}}+3H(\tilde{\rho}+\tilde{p})=0$ and then we obtain the modified 
 Klein-Gordon equation given by
\begin{equation}
	\left[1+2 f(\phi)\right](\ddot{\phi}+3H\dot{\phi})+ f'(\phi)\dot{\phi}^2+\left[1+4 f(\phi)\right]V'+4 f'(\phi)V=0,
\end{equation}
where  the notation $V'$ corresponds to $V'=\frac{\partial V}{\partial \phi}$, $f' = \partial f /\partial \phi$, $V''=\partial^2 V/\partial \phi^2$, etc. 

Also, the Friedmann equation can be written as 
\begin{equation}
H^2=\frac{\kappa}{3}\tilde{\rho}=\frac{\kappa}{3}\,\left(\frac{1}{2}\dot{\phi}^2\left[1+2 f(\phi)\right]+\left[1+4 f(\phi)\right]V\right). \label{HH}
\end{equation}

Following Ref.\cite{1}, we can consider  the slow roll approximation in which $\dot{\phi}^2\ll V$, $\ddot{\phi}\ll H\dot{\phi}$ and  $f'(\phi)\dot{\phi}^2\ll H\dot{\phi}$, respectively. Under this approximation the scalar field and Friedmann equations are reduce to 
\begin{equation}
\label{KGMod}
	3H\dot{\phi}\left[1+2 f(\phi)\right]+\left[1+4 f(\phi)\right]V'+4 f'(\phi)V=3H\dot{\phi}\,F_1+F_2V'+F_2' V\simeq 0,
\end{equation}
and 
\begin{equation}
\label{FriedmannEq}
	H^2\simeq \frac{\kappa}{3}\left[1+4 f(\phi)\right]V=\frac{\kappa}{3}\,F_2\,V, 
 \end{equation}
where we have defined,  the quantities $F_1$ and $F_2$ in terms of the scalar field as 
\begin{equation}
	F_1(\phi)=F_1=1+2f(\phi)\,\,\,\,\mbox{and}\,\,\,
	F_2(\phi)=F_2=2F_1-1=1+4f(\phi).
\end{equation}

On the other hand, in order to give a measure  of the expansion during inflation, we can introduce  the number of $e$-folds $N$ defined from the relation   $\Delta N=N-N_\text{end}=\int^{t_{end}}_t H\,dt=\int^{\phi_{end}}_\phi(H/\dot{\phi})d\phi$, with which using the slow roll approximation yields
\begin{equation}
\Delta N	=\ln[a(t_{end})/a(t)]\simeq\kappa \int^{\phi}_{\phi_{\text{end}}}\left[\frac{(1+2 f(\phi))(1+4 f(\phi))V}{(1+4 f(\phi))V'+4 f'(\phi)V}\right]d\phi,\label{N1}
\end{equation}
where $N$ corresponds to the value of the number of $e$-folds at the cosmological time``t  '' during inflation and  $N_{\text{end}}$ is the $e$-folds at the end of inflation. 

Introducing the dimensionless slow roll parameters, we have \cite{1}
\begin{equation}
	\epsilon_V =\frac{1}{2\kappa (1+2 f(\phi))}\left[\frac{V'}{V}+\frac{4 f'(\phi)}{(1+4 f(\phi))} \right]^2,\label{ep}
\end{equation}
and 
\begin{eqnarray}
\eta_V = \frac{1}{\kappa (1+2 f(\phi))}\left[\frac{V''}{V}+\frac{ f'(\phi)(7+12 f(\phi))}{(1+2 f(\phi))(1+4 f(\phi))}\frac{V'}{V}
-\frac{4 f'(\phi)^2}{(1+2 f(\phi))(1+4 f(\phi))} +\frac{4 f''(\phi)}{(1+4 f(\phi))}\right],\label{et}
\end{eqnarray}
or equivalently in terms of the functions $F_1$ and $F_2$ we get
\begin{eqnarray}
\epsilon_V &=&\frac{1}{2\kappa F_1}\left[\frac{V'}{V}+\frac{F_2'}{F_2} \right]^2,\label{ep1}
\end{eqnarray}
and
\begin{eqnarray}
\eta_V &=& \frac{1}{\kappa F_1}\left[\frac{V''}{V}+\frac{F_1'(1+6F_1)}{2F_1F_2}\frac{V'}{V}
-\frac{F_1'^2}{F_1F_2} +\frac{F_2''}{F_2}\right].\label{et1}
\end{eqnarray} 
The condition under which the inflationary epoch takes place can be summarized with the slow roll parameter $\epsilon_V<1$, which is equivalent to the condition on the scale factor in which $\ddot{a}>0$.  Besides,  we can mention that it is possible to redefine a new field $\tilde{\phi}$ and its effective potential $\tilde{V}$ such  that the matter Lagrangian can be reduced to the canonical form and then the theory can be explicated as standard inflation in the framework of GR from the new scalar field \cite{1}. Thus, using the standard definitions of the slow roll parameters $\epsilon_V$ and $\eta_V$ from the new field $\tilde{\phi}$ and its potential $\tilde{V}$ together with the mapping between $(\tilde{\phi},\tilde{V}) \to (\phi,V)$ one can find the  parameters $\epsilon_V$ and $\eta_V$ given by Eqs.(\ref{ep}) and (\ref{et}), respectively. In this context the relation between these quantities are given by

\begin{equation}
d\tilde{\phi}=(1+2f(\phi))^{1/2}\,d\phi,\,\,\,\,\,\,\mbox{and}\,\,\,\,\,\,\,\tilde{V}=(1+4f(\phi))\,V.\label{Rel}
\end{equation}
Thus, using these relations the slow roll parameters $\epsilon_V$
and $\eta_V$  defined as $\epsilon_V=(1/2\kappa)(\partial\ln\tilde{V}/\partial\tilde{\phi})^2$ and $\eta_V=1/(\kappa\tilde{V})(\partial^2\tilde{V}/\partial\tilde{\phi}^2)$ become the slow roll parameters given by  Eqs.(\ref{ep}) and (\ref{et}), respectively.

In this respect, the scalar  power spectrum of the curvature perturbations $A_s$ and the scalar spectral index $n_s$ defined as $n_s=1+d\ln A_s/d\ln k$, under the slow roll approximation can be written in terms of the slow roll parameters   as \cite{Planck:2013jfk}
\begin{equation}
A_s=\frac{3\kappa\,H^2}{24\pi^2\,\epsilon_V}\simeq\frac{\kappa^2\,(1+4f)\,V}{24\pi^2\,\epsilon_V},\label{AS}
\end{equation}
and
\begin{equation}
	n_s-1=\frac{d\,\ln A_s}{d\,\ln k}\simeq-6\epsilon_V+2\eta_V.	\label{ns}
\end{equation}
In order to demonstrate that from Eq.(\ref{AS}) is possible to find the scalar spectral index given by Eq.(\ref{ns}) in terms of the slow roll parameters $\epsilon_V$ and $\eta_V$ we have
\begin{equation}
n_s-1=\frac{d\,\ln A_s}{d\,\ln k}\simeq\frac{d\,\ln A_s}{dN}=\frac{d\,\ln A_s}{H\,dt}=\frac{\dot{\phi}}{H}\frac{d\,\ln A_s}{d\,\phi}=\frac{\dot{\phi}}{H\,A_s}\frac{d A_s}{d\,\phi},
\end{equation}
where we have considered that $d\ln k\simeq dN$ \cite{Libro2}. In this way, from Eq.(\ref{AS}) we get that the scalar index  can be rewritten as
\begin{equation}
n_s-1=\frac{\dot{\phi}}{H}\,\left[\frac{F_2'}{F_2}+\frac{V'}{V}-\frac{\epsilon_V'}{\epsilon_V}\right]=-2\left[\epsilon_V-\frac{\epsilon_V'}{(V'/V+F_2'/F_2)}\right],\label{ch}
\end{equation}
 where we have used Eqs.(\ref{KGMod}) and (\ref{FriedmannEq}) together with the definition of $\epsilon_V$ given by Eq.(\ref{ep}). Now from Eq.(\ref{et1}) we find that the quantity 
\begin{equation}
 \frac{\epsilon_V'}{(V'/V+F_2'/F_2)}=-2\epsilon_V+\eta_V.\label{ch1}
\end{equation}
Here we have considered the relation $F_2=2F_1-1=1+4f$.
In this form, replacing Eq.(\ref{ch1}) into Eq.(\ref{ch})  we obtain Eq.(\ref{ns}).

 In relation to the generation of the tensor perturbations during the inflationary epoch, we have that the amplitudes of the gravitational waves becomes\cite{Planck:2013jfk}
\begin{equation}
A_t=\frac{2\kappa}{\pi^2}\,H^2\simeq\frac{2\kappa^2\,(1+4f)\,V}{3\pi^2},\label{AT}
\end{equation}
and the tensor to scalar ratio $r=A_t/A_s$ as a function  of the slow roll parameter $\epsilon_V$ from Eqs. (\ref{AS}) and (\ref{AT}) becomes 
\begin{equation}
r\simeq16\,\epsilon_V.\label{r}
\end{equation}

In relation to the tensor spectrum index $n_t$, from the definition $n_t=d\ln A_t/d\ln k$,  we can rewrite  this spectrum index as
\begin{equation}
n_t\simeq\frac{d\ln A_t}{d N}=\frac{\dot{\phi}}{HA_t}\frac{d A_t}{d\phi}=-\left(\frac{F_2'V+F_2V'}{F_2V}\right)^2\frac{1}{\kappa F_1}=-2\epsilon_V,\label{nt}
\end{equation}
where we have used the Eqs.(\ref{KGMod}), (\ref{FriedmannEq}), (\ref{ep1}) and (\ref{AT}), respectively. Thus,  combining  Eqs.(\ref{r}) and (\ref{nt}), we have that the relation between the tensor index and the tensor to scalar ratio or consistency relation becomes $n_t=-r/8$ and it coincides with the obtained in the framework of the GR for a single field  under the slow roll approximation \cite{Planck:2013jfk,Libro2}. In this form, given that the tensor to scalar ratio $r$ is a positive quantity, then the tensor spectrum  index $n_t<0$ and  our model predicts that the tensor spectral index corresponds to a slightly   red-tilted in contradiction to the obtained by the North American Nanohertz Observatory for Gravitational Waves (NANOGrav)\cite{NANOGrav:2020bcs} in which $n_t>0$,  corresponding to a blue-tilted spectrum, see also Refs.\cite{Vagnozzi:2023lwo,Oikonomou:2023qfz}.

In the following, we will study the reconstruction of the effective potential $V(\phi)$ in our inflationary model, by considering as attractor  the spectral index $n_s$ as a function of the number of $e-$folds $N$, i.e., $n_s=n_s(N)$.  Additionally,  we will assume that the coupling function  $f(\phi)$ in terms  of the scalar field $\phi$ is given by 
\begin{equation}
f(\phi)=\alpha_1+\alpha_{2}\,\phi^n,\label{Fp}
\end{equation}
where the $\alpha_1$ and $\alpha_2$ are two constants. The constant  $\alpha_1$ is dimensionless and $\alpha_2$ has dimension of (mass)$^{-n}$. Here we consider that  the power $n$ corresponds to a real number. In the particular case where $\alpha_2=0$ and the constant $\alpha_1\neq 0$ was analyzed in Ref.\cite{Gamonal:2020itt} and 
 recently studied   in Ref.\cite{Ashmita:2022swc} (see also \cite{Chen:2022dyq}). Also, in the situation in which $\alpha_1=0$ and the power $n=1$ together with $\alpha_2\neq 0$, it reduces to the model developed  in Ref.\cite{1}. For the special situation in which $\alpha_1=\alpha_2=0$, then the coupling function $f(\phi)=0$ and the model becomes described by the  GR.

In relation to the coupling function $f(\phi)$ given by Eq.(\ref{Fp}), we mention that 
a   study in  with the 
trace of energy momentum tensor 
 couples to the inflaton field   considering different functions $f(\phi)$ associated to scalar field $\phi$,  such as a simple  expansion of the form $f(\phi)=a_0+a_1\phi+a_2\phi^2+a_3\phi^3+a_4\phi^4+$··· does not
exist in the literature, in order to describe the early and present universe. In particular  in the framework of the reconstruction of the inflation,  the present work  
is the first step towards  that direction using this  coupling function and as  mentioned above different particular cases for the function $f(\phi)$ have been studied in the literature which can be obtained from Eq.(\ref{Fp}).

\section{Reconstruction}

In this section we will study the reconstruction of   the effective potential as a function of the scalar field i.e., $V(\phi)$, assuming as attractor the scalar spectral index $n_s$ in terms of the number of $e-$folds $N$. In this framework considering  the scalar spectral index $n_s(N)$ as a function of $N$, we will first obtain  the effective potential $V$ as a function of the number
of $e-$folds i.e., $V(N)$.

In this way, we need to express the relation between the number of $e-$folds and the scalar field as well 
the scalar spectral index in terms of the number $N$. Thus, using  that
 $V'=\phi_N^{-1}V_N$,  $F_2'=\phi_N^{-1}F_{2\,N}$ and from Eq.(\ref{N1}), we find that the expression for $d\phi/d N=\phi_N$  can be written as
\begin{equation}
\phi_N =\frac{F_2(V'/V)+F'_{2}}{\kappa F_1F_2},\,\,\,\,\,\mbox{and then }\,\,\,\,\,\phi_N^2 =\frac{F_2(V_N/V)+F_{2\,N}}{\kappa F_1F_2}.
	 \label{Nf}
\end{equation}

Here we note that in the particular case in which the coupling function $f(\phi) \to 0$ or analogously $\alpha_1=\alpha_2=0$, the square root of Eq.(\ref{Nf}) with the positive sign of $\phi_N$ reduces to the standard relation given in framework of GR, in which
$\phi_N=\sqrt{\frac{V_N}{\kappa V}}$ \cite{Chiba:2015zpa}. 
In the following,  we will also assume that  the subscription  $V_N$ corresponds to $\partial V/\partial N$, $f_N$ to $\partial f/\partial N $, $\phi_{NN}=\partial^2\phi/\partial N^2$, etc.

Also, we have that the expression for $V''$, $f'$ and $f''$ can be expressed as
\begin{equation}
V''=\frac{V_{NN}}{\phi_{N}^2}-\frac{V_N\,\phi_{NN}}{\phi_N^3},\,\,\,f'=\frac{f_N}{\phi_N}\,\,\,\mbox{and}\,\,\,f''=\frac{f_{NN}}{\phi_{N}^2}-\frac{f_N\,\phi_{NN}}{\phi_N^3},\label{G1}
\end{equation}
respectively.

Besides, we have the relations:
\begin{equation}
(F_2 V)'=\sqrt{\kappa F_1 F_2\,V\,(F_{2N}\,V+F_2\,V_N)},\,\,\,\,\mbox{and}\,\,\,\,(F_1\,V)'=\sqrt{\kappa F_1 F_2\,V\,(F_{1N}\,V+F_1\,V_N)},\label{FP}
\end{equation}
where we have used Eq.(\ref{Nf}).

Analogously, we obtain for the second derivatives
\begin{equation}
(F_2\,V)''=\frac{1}{2(F_2\,V)_N}\left[ F_1\,F_2\,V\,(F_2\,V)_N\right]_N,\,\,\,\,\mbox{and}\,\,\,\,(F_1\,V)''=\frac{1}{2(F_1\,V)_N}\left[ F_1\,F_2\,V\,(F_1\,V)_N\right]_N.\label{FS}
\end{equation}
Note that in the specific case in which the coupling function $f(\phi)=0$ or $F_1=F_2=1$, then  the Eqs.(\ref{FP}) and (\ref{FS}) reduce to the standard form found  for $V'$ and $V''$ in the framework of GR \cite{Chiba:2015zpa}.

From Eq.(\ref{ep}) we find that the slow roll parameter $\epsilon_V$ can be rewritten as
\begin{equation}
\epsilon_V=\frac{1}{2}\,\frac{(F_2\,V)_N}{F_2\,V}=\frac{1}{2}\,\left[\ln(F_2\,V)\right]_N,\label{EP1}
\end{equation}
and the second slow roll parameter $\eta_V$ results
\begin{equation}
\eta_V=\frac{1}{2}\,\left[\frac{(F_2\,V)_N}{F_2\,V}+\frac{(F_2\,V)_{NN}}{(F_2V)_N}\right]=\frac{1}{2}\,\left[\ln(F_2\,V)\right]_N+\frac{1}{2}\,\left[\ln(F_2\,V)_N\right]_N,\label{ET}
\end{equation}
where we have considered Eqs.(\ref{G1}), (\ref{FP}) and (\ref{FS}), respectively.

In this form, combining  Eqs.(\ref{ns}), (\ref{EP1}) and (\ref{ET}), we get that the scalar spectral index can be rewritten as
\begin{equation}
	n_s-1=\left(\text{ln}\left[\frac{V_N F_2+V\,F_{2_N})}{V^2\,F_2^2} \right]\right)_N=\left(\text{ln}\left[\frac{V_N (1+4f)+4V\,f_N)}{V^2(1+4f)^2} \right]\right)_N.
\label{ns2}
\end{equation}
By using Eq.(\ref{ns2}) we can obtain  a first integral yields 
\begin{equation}
\frac{1}{V^2(1+4f)^2}\frac{d}{dN}[V(1+4f)]=\exp\left[\int (n_s-1)dN\right].
\end{equation}

In this way, using the above equation  we get an expression for  $\tilde{V}=V(1+4f)$ in terms of the number of $e$-folds 
given by 
\begin{equation}
\tilde{V}(N)=\,\left[-\int \left(\exp\left[\int (n_s-1)dN\right]\right)dN\right]^{-1}.\label{V1}
\end{equation}

From Eq.(\ref{Nf}) we find that the relation between the scalar field and the number of $e$-folds  can be obtained from the relation
\begin{equation}
\phi_N=\sqrt{\frac{(F_2\,V)_N}{\kappa\,F_1\,F_2\,V}}=\sqrt{\frac{\tilde{V}_N}{\kappa\,F_1\,\tilde{V}}}\,\,.\label{F1a}
\end{equation}

The equations expressed  by Eqs.(\ref{V1}) and (\ref{F1a}) together with the coupling function $f(\phi)$ given by Eq.(\ref{Fp}) constitute  the basic
equations for reconstructing the scalar potential  as a function of the inflaton field i.e., $V(\phi)$.

In the following,
we will consider a particular example in order to rebuild
this background variable from the parameterization of the
scalar spectral index $n_s$ in terms of  the number of $e$-folds $N$ i.e., $n_s=n_s(N)$, in which we will utilize a specific attractor  $n_s(N)$.

\section{  Attractor: An Example for the scalar spectral index $n_s(N)$}
 
In order to apply the method of above, we shall assume the simplest ansatz for  the observational parameter $n_s(N)$  to find the reconstruction of the effective potential as a function of the scalar field i.e., $V(\phi)$.  In this context, we can consider the simplest attractor for the scalar spectral index $n_s(N)$   given by \cite{Chiba:2015zpa} \begin{equation}
	n_s(N)=1-\frac{2}{N},\,\,\,\,\;\;\mbox{with}\,\,\,\;\;\;N\neq 0.\label{ns1}
\end{equation}
The attractor given by Eq.(\ref{ns1}) corresponds to a parameterization of $n_s$ in term of the number $N$ for large $N$. Here   large $N$ signifies that the number of $e$-folds $N\sim\mathcal{O}(10)\sim\mathcal{O}(10^2)$, throughout the slow roll regime during the inflationary era \cite{Chiba:2015zpa}. In addition,  we mention that the parameterization of the scalar spectral index as a function of the number of $e-$ folds $N$ given by Eq.(\ref{ns1}) arises of  the hyperbolic tangent model or T-model in the framework of the GR studied in Ref.\cite{M1}.

By replacing Eq.(\ref{ns1}) into Eq.(\ref{V1}) we obtain that  
\begin{equation}
	\tilde{V}(N)=\left[\frac{A}{N}+B\right]^{-1},\label{VVV}
\end{equation}
where $A$ and $B$ correspond to two integration constants. The integration constant $A$ is positive from the Eq.(\ref{Nf}) in which we have chosen the positive sign, however the constant $B\lesseqqgtr 0$.

In the situation in which the integration constant  $B>0$ and defining the quantity $\mu^2=B/A$, then from Eqs.(\ref{Fp}) and (\ref{Nf}) we find that the relation between the number $N$ and the scalar field can be written as 

\begin{equation}
	N(\phi)=\frac{1}{\mu^2}\text{sinh}^2\left(\frac{\mu\sqrt{\kappa}}{2}F_3(\phi)\,+C_1\right),\label{NNN}
\end{equation}
where the function $F_3(\phi)$ is defined as
\begin{equation}
\label{F3}
   F_3(\phi)= \frac{\left(1+2\alpha_1+2\alpha_2\phi^n \right)^{3/2}\phi}{(1+2\alpha_1)}\,\,\,_{2}F_{1}\left[1,\frac{3}{2}+\frac{1}{n},1+\frac{1}{n},-\frac{2\alpha_2\phi^n}{(1+2\alpha_1)}\right],
\end{equation}
and $C_1$ is a new integration constant. Also, the quantity $_{2}F_1$ corresponds to the hypergeometric function \cite{Libro}. When the power $n=0$ i.e., the function $f(\phi)=$cte., then the hypergeometric function reduces to $_2F_1\propto$ cte , $F_3(\phi)\propto \phi$ and the number of $e$-folds $N\propto \text{sinh}^2[\mbox{cte}\,\phi]$ as the standard case \cite{Chiba:2015zpa}. For $n=1$ the term $\phi\,\,\,_2F_1\propto \phi^{3/2}$, etc.

Note that from Eq.(\ref{F1a}) we can obtain an expression for the scalar field in terms of the number of $e$-folds $N$ given by  \cite{Libro}
\begin{equation}
\phi=\mathcal{F}^{-1}(N),\label{PF1}
\end{equation}
where the function $\mathcal{F}^{-1}(N)$ represents the inverse function of 
\begin{equation}
{\mathcal{F}}(N)=\sqrt{\kappa(1+2\alpha_1)}\,\,\,g(N)\,\,\,_2F_1\left[-\frac{1}{2},\frac{1}{n},1+\frac{1}{n},-\frac{2\alpha_2}{(1+2\alpha_1)}\,g(N)^n\right],
\end{equation}
in which the function $g(N)$ is defined as
\begin{equation}
g(N)=\left(\frac{2}{\mu}\right)\mbox{arcsinh}[\mu\,\sqrt{N}]+C,
\end{equation}
where $C$ corresponds to an integration constant.

Now replacing Eq.(\ref{Fp}) together with the number of $e$-folds $N$ given by Eq.(\ref{NNN})   into Eq.(\ref{VVV}), we obtain that the reconstruction from the parameterization given by Eq.(\ref{ns1}) for the effective potential becomes

\begin{equation}
	V(\phi)=\frac{1}{B[1+4(\alpha_1+\alpha_2\,\phi^n)]}\text{tanh}^2\left(\frac{\mu\sqrt{\kappa}}{2}\,F_3(\phi)+C_1
	\right).\label{Vff}
\end{equation}

\begin{figure}
    \centering
    \includegraphics[width=9
cm]{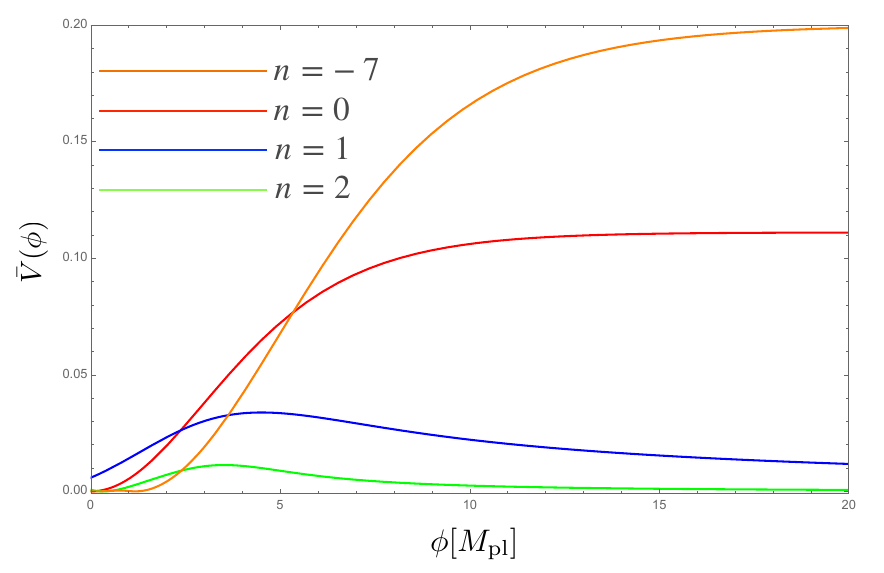}
    \caption{The figure shows the evolution of the dimensionless effective potential   $\bar{V}(\phi)=BV(\phi)$ versus the scalar field $\phi$ from Eq.(\ref{Vff}), for four values of the power $n$ related to the coupling function $f(\phi)$. Here we have considered the values $\alpha_1=1$, $\alpha_2=\kappa^{n/2}$, $\mu=0.2$ and the integration constant $C_1=0$.}
    \label{fig1a}
\end{figure}

As before for the special case in which $n=0$, we have that the reconstruction of the effective potential $V(\phi)\propto \text{tanh}^2[\mbox{cte}\,\phi] $ and this potential corresponds to the T-model inflation studied in  \cite{M1}. For the situation with $n=1$, the potential reduces to $V(\phi)\propto \text{tanh}^2[\mbox{cte}\,\phi^{3/2}]/\phi $, etc.

In Fig.\ref{fig1a} we show the evolution of the dimensionless effective potential $\bar{V}(\phi)=BV(\phi)$ on the scalar field $\phi$ and as it increases in terms of the scalar field to later approach to a constant value for large-$\phi$ ($>$10 M$_\text{pl}$).  In the plot we have used Eq.(\ref{Vff}) where the constant $B>0$, together with different values of the power $n$  associated to the coupling function $f(\phi)$ given by Eq.(\ref{Fp}).  
From the plot we note that if we increase the value of the power $n$ to positive values, the reconstructed  effective potential results to be smaller. 
 Besides, from Fig.\ref{fig1a} we note that  for the specific case in which $n=1$, the reconstructed potential $V(\phi)$ shows a maximum value around   $\phi\simeq 5M_{pl}$ and then it decreases for large-$\phi$. However, from Eq.(\ref{NNN}) it is possible to find numerically that the value of the scalar field at $N=60$  corresponds to $\phi(N=60)\simeq 4.5M_{pl}$.  Thus, this suggests that  the scalar field begins to roll from the maximum value of the potential  towards values of $\phi\sim 0$ in which $N_{end}\sim 0$, during the inflationary scenario. 

In this context, in the cases in which the power $n>0$, we have that  the first function  associated to the effective potential (\ref{Vff})  given by $1/[1+4(\alpha_1+\alpha_2\phi^n)]$ tends to zero for very large $\phi$ and then this function  predominates over the second  function $\tanh^2[\mu\sqrt{\kappa}\,F_3(\phi)/2]$ which becomes a constant for large-$\phi$, with which the potential $V\rightarrow 0$ from a maximum value of $V$, see  Fig.\ref{fig1a}  the cases $n=1$ and $n=2$. In this sense, from Eq.(\ref{F1a}) we have assumed that $\phi_N$ is a positive quantity and in the same form $V_N>0$, in order to reproduce the GR when $\alpha_1=\alpha_2=0$ ($\tilde{V}_N>0$), then from the definition  $V'=\phi_N^{-1}V_N$ we have that  $V'$ is a positive quantity. Thus,  in the framework of the reconstruction of the effective potential in terms of the scalar field only  makes sense  when $V'>0$.  In this form,  we can restrict   the evolution of the scalar field for the special cases in which the power $n$ is positive. In this context, we can constrain the evolution of the scalar field during the inflationary scenario, such that 
the  range for the field $\phi$  is given by  $0<\phi\lesssim\phi_{max}$, where the quantity  $\phi_{max}$ corresponds to the maximum value of the scalar field associated to the maximum value of the potential $V(\phi)$. Here  the maximum value of the scalar field can be obtained  by the condition $V'(\phi=\phi_{max})=0.$ Additionally, 
we know that the energy density of the universe  associated to scalar field cannot increase over time, then    the field cannot reach the maximum of the potential from the right (large-$\phi$) in which $V'<0$. 

 Also,  we note that the relation between the redefined field $\tilde{\phi}$ and the number of $e-$folds $N$ from Eqs.(\ref{Rel}) and (\ref{NNN}) is given
$N(\tilde{\phi})=\mu^{-2}\text{sinh}^2\left(\mu\sqrt{\kappa}\tilde{\phi}/2\,+C_1\right),
$
since $\tilde{\phi}=F_3(\phi)+Cte$. In this way, replacing 
 this number of $e-$folds into Eq.(\ref{VVV}), we find that the scalar potential in terms of the redefined field is reduced to the potential obtained in the framework of the standard GR (T-model)\cite{M1} yields 
 $\tilde{V}(\tilde{\phi})=B^{-1}\,\tanh^2\left(\mu\sqrt{\kappa}\,\tilde{\phi}/2+C_1\right),
 $
and   this potential does not depend on the power $n$ and then the potential $\tilde{V}(\tilde{\phi})$ does not present the behavior of the potential (\ref{Vff}) for large-$\phi$ when the power  $n>0$.

For the case in which the integration constant $B<0$, we have that the parameter $\mu$ is imaginary i.e., $\mu=i\sqrt{\lvert B \rvert/A}$ and then the reconstructed effective potential can be written as
\begin{equation}
	V(\phi)=-\frac{1}{B[1+4(\alpha_1+\alpha_2\,\phi^n)]}\text{tan}^2\left(\sqrt{\frac{\kappa \rvert B\rvert}{4A}}\,F_3(\phi)+C_1
	\right).\label{Vff2}
\end{equation}
In the situation in which the integration constant $B=0$, we get that the reconstruction of the effective potential becomes
\begin{equation}
	V(\phi)=\frac{\kappa}{4A[1+4(\alpha_1+\alpha_2\,\phi^n)]}\left(F_3(\phi)+C_1
	\right)^2.\label{Vff2}
\end{equation}
In particular for $n=0$ the reconstructed potential given by Eq.(\ref{Vff2}) reduces to the quadratic chaotic potential in which $V(\phi)\propto \phi^2$.

On the other hand, in order to obtain a constraint on the integration constant $A$,  we can consider the scalar power spectrum given by Eq.(\ref{AS}) such that
\begin{equation}
	A_s=\left.\frac{3\kappa \,H^2}{24\pi^2\,\epsilon_V}\right|_{k=a_kH_k}= \left.\frac{\kappa^2}{12\pi^2}
\frac{N^2}{A}\right|_{k=a_kH_k}=\frac{\kappa^2}{12\pi^2}
\frac{N_k^2}{A},
\end{equation}
and then  we find that the integration constant $A$ becomes
\begin{equation}
    A=\frac{\kappa^2}{12\pi^2}\label{AA}
\frac{N_k^2}{A_s}.
\end{equation}

Additionally, in the specific case in which the integration constant $B>0$, we can determine a constraint for $B$ from the tensor to scalar ratio defined  by $r=16\epsilon_V$. In this context, we find that the tensor to scalar ratio $r$ as a function of the number of $e$-folds when the universe scale crosses the Hubble horizon becomes
\begin{equation}
    r(k)|_{k=a_kH_k}=r_k=\left.16\,\epsilon_V\right|_{k=a_kH_k}=\left.\frac{8}{(N+\mu^2 \,N^2)}\right|_{k=a_kH_k}=\frac{8}{(N_k+\mu^2 \,N_k^2)},
\label{R2}
\end{equation}
where we have assumed  that the tensor to scalar ratio $r(k)|_{k=a_kH_k}=r_k$ corresponds to the ratio when the universe scale crosses the Hubble horizon during the inflation and its upper bound is $0.039$ at $k=0.05$ Mpc$^{-1}$ ($96\%$ confidence level). In this way,  from Eq.(\ref{R2}) we can obtain that a lower bound for the integration constant $B$  given by 
\begin{equation}
	B>\frac{A}{N_k}\left(\frac{8}{N_k\,r_k}-1 \right),\label{BB}
\end{equation}
where in order to satisfy the condition in which  the constant $B>0$, we impose the restriction  $8>N_k\,r_k$. In this sense,  considering the particular case in which  the number of $e$-folds during the crossing $N_k=60$, and the cosmological parameters $A_s=2.2\times 10^{-9}$ and the tensor to scalar ratio $r_k=0.039$, we find that the integration constants are given by 
$$
A\simeq1.4\times 10^{10}\kappa^2,\,\,\,\, \mbox{and}\,\,\, B\,>\,5.6\times10^{8}\kappa^2,
$$
respectively. Here we have used Eqs.(\ref{AA}) and (\ref{BB}), respectively.

On the other hand, in order to calculate the value of the number of $e$-folds at the end of inflationary epoch $N_\text{end}$ in our model, we can consider from 
 Eq.(\ref{EP1})  the parameter $\epsilon_V$ as a function of the number $N$, resulting in
\begin{equation}
    	\epsilon_V(N)=\frac{1}{2}\left[\ln(F_2\,V)\right]_N=\frac{1}{2N\,(1+\mu^2\,N)}.
\end{equation}
Now, by using the fact that inflation ends when $\epsilon_V(N=N_{end})=1$, we obtain that the  solution for the number of $e$-folds at the end of inflation becomes 
\begin{equation}
\label{Nend2}
   N_\text{end}=\frac{\sqrt{1+2\mu^2}-1}{2\mu^2}.
\end{equation}
Note that for $1\gg \mu^2$ the number of $e-$ folds at the end of inflation is  $N_{end}\simeq 0$. In particular using the values of the integration constants $A=1.4\times 10^{10}M_{pl}^{-4}$ and the lower bound for $B=5.6\times 10^{8}M_{pl}^{-4}$, we obtain that the dimensionless parameter $\mu^2=B/A\simeq 0.04$ and then the number $N_\text{end}\simeq 0.5$.

In the following we will analyze the reheating of our model from the reconstructed effective potential assuming the attractor $n_s-1=-2/N$. Besides, during this scenario we shall determine the reheating temperature together with the number of $e$-folds during this regime.

\section{ Reheating era}

In section we will analyze the   reheating scenario for our model by considering the reconstruction of the effective potential in terms of the scalar field obtained in the previous  section. For simplicity in the following we will consider the effective potential given by Eq.(\ref{Vff}) in which the integration constant $B$ is positive.

In order to study the reheating epoch, we can consider that the ratio between the comoving Hubble scale crosses the horizon during the inflationary epoch is $k=a_k H_k$ and the scale $k_0$ which equals the present Hubble scale corresponds to $k_0=a_0H_0$ \cite{Liddle:2003as}
\begin{equation}
	\frac{k}{k_0}=\frac{a_kH_k}{a_0H_0}=\left(\frac{a_k}{a_\text{end}}\right)\left(\frac{a_\text{end}}{a_\text{reh}}\right)\left(\frac{a_\text{reh}}{a_\text{eq}}\right)\left(\frac{a_\text{eq}H_\text{eq}}{a_0H_0}\right)\left(\frac{H_k}{H_\text{eq}}\right),\label{kk}
\end{equation}
where the different quantities with the subscript ``end'' refer to the end of inflation, ``reh'' correspond to the reheating and ``eq'' denote the radiation-matter equality. As before, we recalled the quantities with subscript $k$  are evaluated at horizon exit during the inflationary era.

By writing the number of $e$-folds $N$ in terms of the scale factor $a$ during the different epochs we get    
\begin{equation}
	e^{N_k}=\frac{a_\text{end}}{a_k},\,\,\,\,\,\,\,\,\,
	e^{N_\text{reh}}=\frac{a_\text{reh}}{a_\text{end}},\,\,\,\,\,\,\,\,
\mbox{and}\,\,\,\,\,\,\,\,\,\,\,\,	e^{N_{RD}}=\frac{a_\text{eq}}{a_\text{reh}},\label{ar}
\end{equation} 
where  $N_\text{reh}$ corresponds to the number of $e$-folds during reheating regime and $N_\text{RD}$ is the number of $e$-folds during radiation dominance. Thus, from Eq.(\ref{kk}) we have 
\begin{equation}
\label{NewNreh}
	\text{ln}\left(\frac{k}{a_0H_0}\right)=-N_k-N_\text{reh}-N_\text{RD}+\text{ln}\left(\frac{a_\text{eq}H_\text{eq}}{a_0H_0}\right)+\text{ln}\left(\frac{H_k}{H_\text{eq}}\right).
\end{equation}
On the other hand, from Eq.(\ref{ar}) and considering that the energy density at the end of inflationary epoch $\rho_\text{end}$ and the energy density at the end of the reheating regime $\rho_\text{reh}$ are related from  an Equation of State (EoS) parameter  $w_\text{reh}$ associated to the reheating regime,    we find  that the ratio $\rho_\text{reh}/\rho_\text{end}$ results
\begin{equation}
	\frac{\rho_\text{reh}}{\rho_\text{end}}=e^{-3N_\text{reh}(1+w_\text{reh})}.
\end{equation}
Here we have used that during the reheating regime the energy density decays in terms of the scale factor as
 $\rho\propto a^{-3(1+w_\text{reh})}$, where the EoS parameter $w_\text{reh}=$ constant.

On the other hand, from Eq.(\ref{HH}) we find that  the energy density at the end of inflationary epoch  is given by 
\begin{equation}
	\rho_{\text{end}}=[1+4f(\phi_\text{end})]V_{\text{end}}\left(1+\frac{\dot{\phi}^2_\text{end}}{2 V_{\text{end}}}\frac{[1+2f(\phi_\text{end})]}{[1+4f(\phi_\text{end})]} \right)
                     =[1+4f(\phi_\text{end})]V_{\text{end}}(1+\lambda),
\label{RHE}
\end{equation}
where the quantity $\lambda$  is defined at the end of inflation  as $\lambda=\frac{\dot{\phi}^2_\text{end}}{2 \tilde{V}_{\text{end}}}\frac{[1+2f(\phi_\text{end})]}{[1+4f(\phi_\text{end})]}$. 

By using the  definition of the slow-roll  parameter $\epsilon$ in terms of the Hubble parameter yields 
\begin{equation}
    \epsilon =-\frac{\dot{H}}{H^2}
             \simeq \frac{3\dot{\phi}^2(1+2f)/2V(1+4f)}{1+\dot{\phi}^2(1+2f)/2V(1+4f)}.
\end{equation}
In this way,  we find that the quantity $\lambda$ which is  evaluated at the end of inflation can be rewritten as a function of the slow roll parameter at the end of inflation $\epsilon_\text{end}$ as 
\begin{equation}
  \lambda =\frac{\epsilon_\text{end}}{3-\epsilon_\text{end}},  
\end{equation}
and considering that inflation ends when the slow roll parameter $\epsilon_\text{end}=1$ (or equivalently $\ddot{a}=0$), then we have that $\lambda \simeq 1/2$ and from Eq.(\ref{RHE}) we get $\rho_{\text{end}}\simeq (3/2) [1+4f(\phi_\text{end})]V_{\text{end}}$. 
Thus, using Eq.(\ref{Vff}) we get that the effective potential at the end of inflation for the case in which the integration constant $B>0$, resulting in   
\begin{equation}
	V(\phi=\phi_\text{end})=V_{\text{end}}=\frac{1}{B[1+4(\alpha_1+\alpha_2\,\phi^n_\text{end})]}\text{tanh}^2\left(\frac{\mu\sqrt{\kappa}}{2}\,F_3(\phi_\text{end})+C_1
	\right),\label{Vfff}
\end{equation}
and then the energy density $\rho_\text{end}$ is given by 
\begin{equation}
\label{RhoEnd}
	\rho_\text{end}\simeq\frac{3}{2B}\text{tanh}^2\left(\frac{\mu\sqrt{\kappa}}{2}\,F_3(\phi_\text{end})+C_1
	\right).
\end{equation}

On the other hand, following Ref.\cite{Dai:2014jja} we can consider the entropy conservation in which the reheating entropy is preserved in the CMB together with the  neutrino background today in order to find a relation between the reheating temperature $T_\text{reh}$ and the scale factor $a_\text{reh}$. In this context, we have \cite{Dai:2014jja} 
\begin{equation}
	g_{\text{s,reh}}a^3_{\text{reh}}T^3_{\text{reh}}=a_0^3\left(2T_0^3+\frac{21}{4}T_{\nu 0}^3\right),
\end{equation}
where $g_{\text{s,reh}}$ corresponds to the effective number of relativistic degrees of freedom for entropy at reheating, $T_{\nu 0}$ is the present neutrino temperature and the current CMB temperature $T_0\simeq2.7$K. Thus,  considering that the relation between the neutrino temperature $T_{\nu 0}$ and $T_0$ is given by 
 $T_{\nu 0}=\left(\frac{4}{11}\right)^{1/3}T_0$ \cite{Dai:2014jja}, we can relate the scale factors during the different epochs as 
\begin{equation}
	\frac{a_{\text{reh}}}{a_0}=\left(\frac{43}{11g_{\text{s,reh}}}\right)^{1/3}\frac{T_0}{T_{\text{reh}}}.
\end{equation}
 Additionally, we can consider  that the energy density at the end of reheating is equivalent to the hot radiation with which
\begin{equation}
\label{Treh1}
	\rho_{\text{reh}}=\frac{\pi^2}{30}g_{\star\text{,reh}}T^4_{\text{reh}},
\end{equation}
where $g_{\star\text{,reh}}$ denotes  the effective number of relativistic degrees of freedom at the end of reheating. By combining  the above equations,  we find that $T_\text{reh}$ can be written as
\begin{equation}
\label{TrehVend}
	T_\text{reh}=\text{exp}\left[-\frac{3}{4}(1+\omega_\text{reh})N_\text{reh}\right]\left(\frac{9}{20B\pi^2}\right)^{1/4}\text{tanh}^{1/2}\left(\frac{\mu\sqrt{\kappa}}{2}F_3(\phi_\text{end})\,+C_1
	\right),
\end{equation}
where the number of $e$-folds during the reheating regime becomes 
\begin{equation}
N_\text{reh} =\frac{4}{1-3\omega_\text{reh}}\left[- N_k-\text{ln}\left(\frac{ k}{a_0T_0}\right)  -\frac{1}{3}\text{ln} \left(\frac{11g_\text{s,reh}}{43}\right)
	-\frac{1}{4}\text{ln}\left(\frac{30 \kappa^2\rho_\text{end}}{g_\text{reh}\pi^2}\right)+\frac{1}{2}\text{ln}\left(\frac{\pi^2 r A_s}{2}\right)\right].
\end{equation}
Here we have considered that the Hubble parameter $H_k$ can be written in terms of scalar-tensor ratio $r$ and the scalar power spectrum  $A_s$ through  
$H_k=\sqrt{\frac{\pi^2}{2\kappa}rA_s}$.

Also,  we can rewrite the reheating temperature and the number of $e$-folds during the reheating stage in terms of the scalar spectral index $n_s$ as
\begin{equation}
	T_\text{reh}(n_s)=\left[\frac{9\left(\sqrt{1+2\mu^2}-1\right)}{20\pi^2\,A\,\mu^2[1+4(\alpha_1+\alpha_2\phi^n_\text{end}(n_s))]\left(1+\sqrt{1+2\mu^2}\right)}\right]^{1/4}\text{exp}\left[-\frac{3}{4}(1+\omega_\text{reh})N_\text{reh}(n_s)\right],
\label{I1}
\end{equation}
and 
\begin{equation}
N_\text{reh}(n_s) =\frac{4}{1-3\omega_\text{reh}}\left[ 2(n_s-1)^{-1}-\text{ln}\left(\frac{ k}{a_0T_0}\right)  -\frac{1}{3}\text{ln} \left(\frac{11g_\text{s,reh}}{43}\right)
	-\frac{1}{4}\text{ln}\left(\frac{30 \kappa^2\rho_\text{end}(n_s)}{g_\text{reh}\pi^2}\right)+\frac{1}{2}\text{ln}\left(\frac{\pi^2 r(n_s) A_s(n_s)}{2}\right)\right],
\label{I2}
\end{equation}
respectively. 

Here the energy density at the end of inflation as a function of $n_s$ yields
\begin{equation}
\rho_\text{end}(n_s)=\frac{3\left(\sqrt{1+2\mu^2}-1\right)}{2\,A\,\mu^2[1+4(\alpha_1+\alpha_2\phi^n_\text{end}(n_s))]\left(1+\sqrt{1+2\mu^2}\right)},
\end{equation}
and the tensor to scalar ratio and the scalar power spectrum as a function of the scalar spectral index become 
\begin{equation}
r(n_s)=\frac{4(1-n_s)^2}{(1-n_s+2\mu^2)},\,\,\,\,\,\mbox{and}\,\,\,\,\,\;\;A_s(n_s)=
\frac{\kappa^2}{3\pi^2\,A\,(1-n_s)^2}.
\end{equation}
Additionally, we have that the scalar field $\phi_\text{end}=\phi_\text{end}(n_s)$ as a function of the scalar spectral index $n_s$, is given by Eq.(\ref{PF1}) in which $\phi_\text{end}(n_s)=\mathcal{F}^{-1}(2/(1-n_s))$. 

By considering that the energy density at the end of reheating stage should be smaller that the energy density associated to the end of the inflationary epoch, we can consider an upper limit on the reheating temperature or called critical reheating temperature $T^c_\text{reh}$ such that

\begin{equation}
T^{\text{c}}_{\text{reh}}\simeq2.4\times10^{18}\ \left(\frac{135A_s}{2g_\text{reh}\mu^4}\right)^{1/4}\left[(1-n_s)(\sqrt{1+2\mu^2}-1)\right]^{1/2}\text{GeV}=2.9\times 10^{18}\left[\frac{\kappa^2(\sqrt{1+2\mu^2}-1)^2}{A\,g_\text{reh}\mu^4}\right]^{1/4}\text{GeV},
\end{equation}
where we have considered Eq.(\ref{Nend2}) and the relation
$  \frac{\pi^2}{30}g_{\text{reh}}T^4_{\text{reh}}\lesssim \frac{3}{A}N_\text{end}^2.
$

\begin{figure}
\includegraphics[width=18cm]{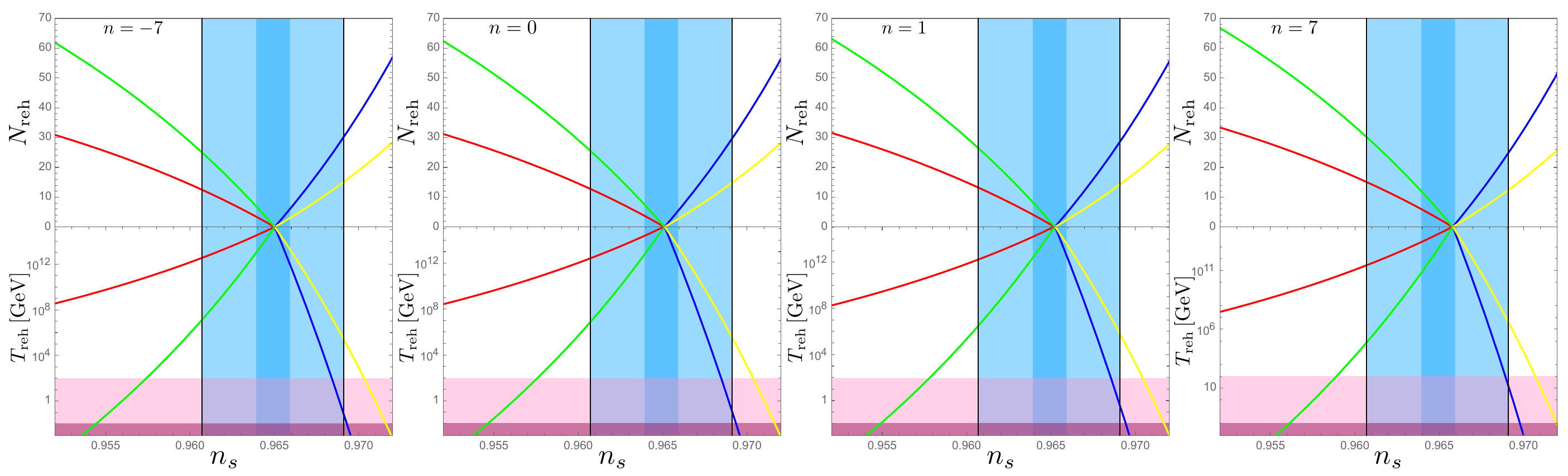}
\caption{ The plots show the number of $e$-folds (upper panels) during the reheating and the reheating temperature (lower panels) versus the scalar spectral index $n_s$ for different power of $n$ associated to the coupling function $f(\phi)$. Here  we have fixed that the constants $\alpha_1$ and $\alpha_2$ are given by  $\alpha_1=1$ and $\alpha_2=1\times M_\text{pl}^{-n}$, respectively. In each panel, we have considered various values for the effective  EoS parameters $w_\text{reh}$. From the left to right  we have considered that;  the EoS parameter $w_\text{reh}=-1/3$ corresponds to red  line, $w_\text{reh}=0$ to green line, $w_\text{reh}=2/3$ to blue line and  $w_\text{reh}=1$ to yellow line, respectively.
The light blue shaded region
corresponds to 
the maximum and minimum values of the scalar spectral index $n_s=0.9649\pm 0.0042$ from data Planck (1$ \sigma$ bounds on $n_s$). The blue shaded indicates a projected sensitivity in the experiment of $\pm 10^{-3}$ (see e.g. \cite{-3}) from the central value $n_s=0.9649$. Additionally, 
the pink shaded region is  below the electroweak scale in which the temperature becomes $T_\text{EW}\sim100\text{GeV}$ and the purple shaded region with temperatures below   10 MeV,  and this region is discarded by BBN. 
  }
   \label{fig1}
\end{figure}

In Fig.\ref{fig1} we show the results above to determine the number of $e$-folds  and the temperature  during the reheating scenario i.e., $N_\text{reh}$ and $T_\text{reh}$  as functions of the scalar spectral index $n_s$. Here we have considered different values for the power $n$ associated to the coupling function $f(\phi)$ defined by Eq.(\ref{Fp}). In this plot we have fixed the dimensionless constant  $\alpha_1=1$ and the constant $\alpha_2$ of dimension of (mass)$^{-n}$ to $\alpha_2=1\times \text{M}_\text{pl}^{-n}$ (if $n=0$ then $\alpha_2=1$, $n=1$ corresponds to  $\alpha_2=1/M_\text{pl}$, etc.). 
In particular in these panels  we analyze from left to right  the coupling  function for the powers $n=-7$, $n=0$, $n=1$ and $n=7$, respectively. Additionally, different EoS parameters $w_\text{reh}$ associated to the reheating era are assumed in each situation.  Further, the scenario of  instantaneous reheating is defined when the number of $e$-folds $N_\text{reh}$ tends to zero (at the end of inflationary epoch)  and it corresponds   to the specific point where all the lines converge as we see in the plots. Also, during the instantaneous reheating the model archives the maximum temperature and it occurs at the end of reheating in which $N_\text{reh}\sim 0$. Also, we can note that during the instantaneous reheating  the EoS parameter $w_\text{reh}$ is irrelevant, since all curves associated to the EoS parameters converge to the same point. 

Besides, from Fig.\ref{fig1} we show that in the specific case in which the constants $\alpha_1$ and $\alpha_2$ are similar and close to unity, the model has a very good compatibility with Planck's 1$\sigma$ bounds on the scalar spectral index $n_s$, for  the different values of the EoS parameters $w_\text{reh}$, excepting  the EoS parameter $w_\text{reh}=-1/3$ (red line) when the temperature $T_\text{reh}<10^{13}$GeV. Even more, for values of the EoS parameter $w_\text{reh}\ge 0$, we can observe that this compatibility with the Planck-data becomes independent of the power $n$ (from $n=-7$ to $n=7$) associated to the coupling function $f(\phi)$. Also, we have that any temperature between the BBN limit and the instantaneous reheating value is  permitted inside the Planck's 1$\sigma$ limit regardless of the values of EoS parameter $w_\text{reh}$ and the power $n$ excepting the specific value $w_\text{reh}=-1/3$. In this respect, from the Planck data (1$\sigma$ bound on $n_s$) we can observe that the model for the cases in which the EoS parameter  $w_\text{reh}\ge 0$ predicts higher values related to the reheating temperatures, in particular for $w_\text{reh}=0$ with $T_\text{reh}\gtrsim 5\times10^{7}$GeV. 
In relation to the number of $e$-folds during the reheating epoch, we note that in this situation the model predicts a small $N_\text{reh}\lesssim 30$ for different reheating temperatures.

\begin{figure}
    \centering
    \includegraphics[width=18cm]{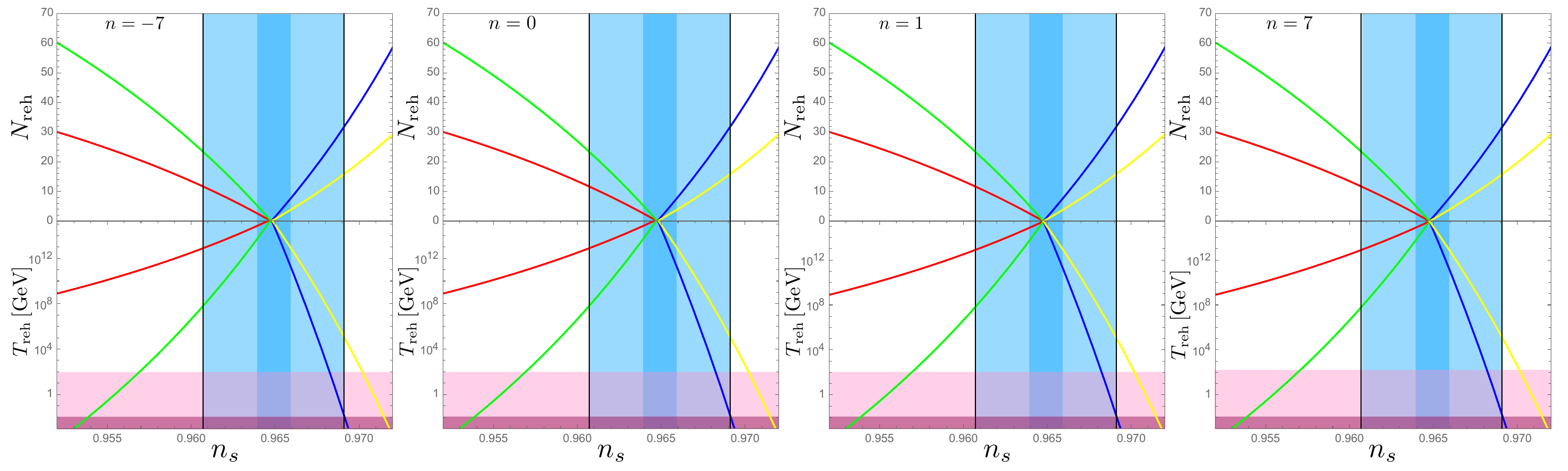}
    \caption{The plots show the number $N_\text{reh}$
    and the temperature $T_\text{reh}$ versus the scalar spectral index $n_s$, for the case in which both parameters $\alpha_1$ and $\alpha_2\kappa^{-n/2}$ are much smaller than unity, i.e.,  $(\alpha_1,\alpha_2\kappa^{-n/2})\ll 1$. In all plots the curves and shaded regions  are as the Fig.(\ref{fig1}).  In particular here we have used for $\alpha_1$ and $\alpha_2$ the values $\alpha_1=10^{-15}$ and $\alpha_2=1\times 10^{-15}M^{-n}_\text{pl}$, respectively.
    }
    \label{fig2}
\end{figure}

\begin{figure}
    \centering
    \includegraphics[width=18cm]{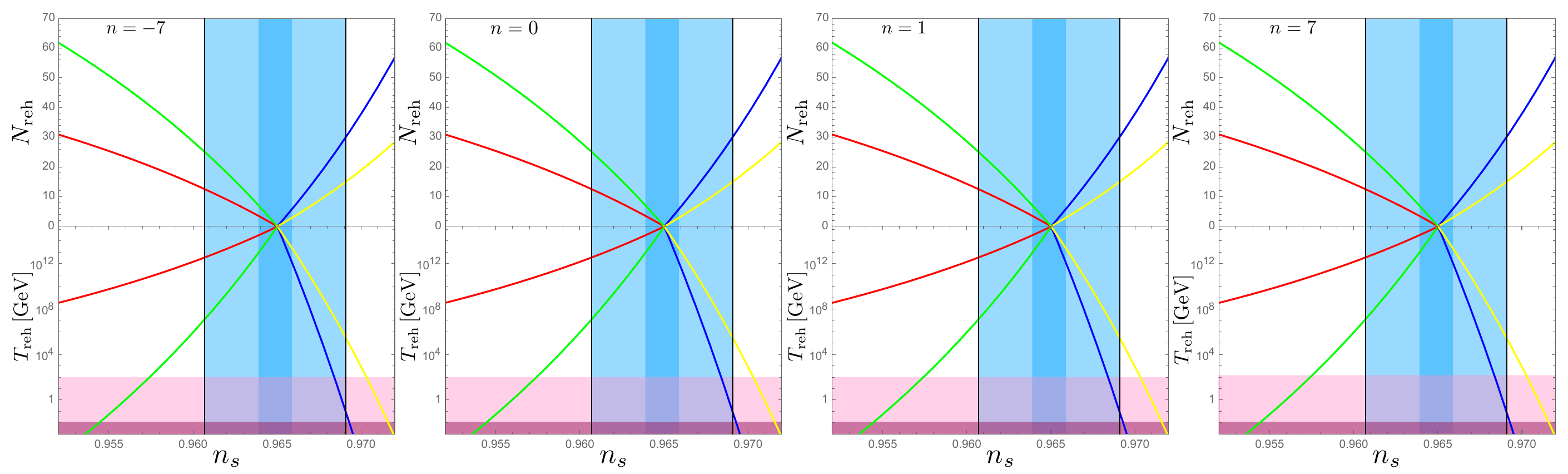}
    \caption{The plots show the length of reheating era from $N_\text{reh}$
    and the reheating temperature $T_\text{reh}$ versus the scalar spectral index $n_s$ for the model, in the situation  in which the parameter $\alpha_1\sim\mathcal{O}(1)$ and $\alpha_2\kappa^{-n/2}\ll 1$ . As before in all plots the curves and shaded regions  are as the Fig.(\ref{fig1}).  Here we have used for $\alpha_1$ and $\alpha_2$ the values $\alpha_1=1$ and $\alpha_2=1\times 10^{-15}M^{-n}_\text{pl}$, respectively.
}
    \label{fig3}
\end{figure}

\begin{figure}
    \centering
    \includegraphics[width=18cm]{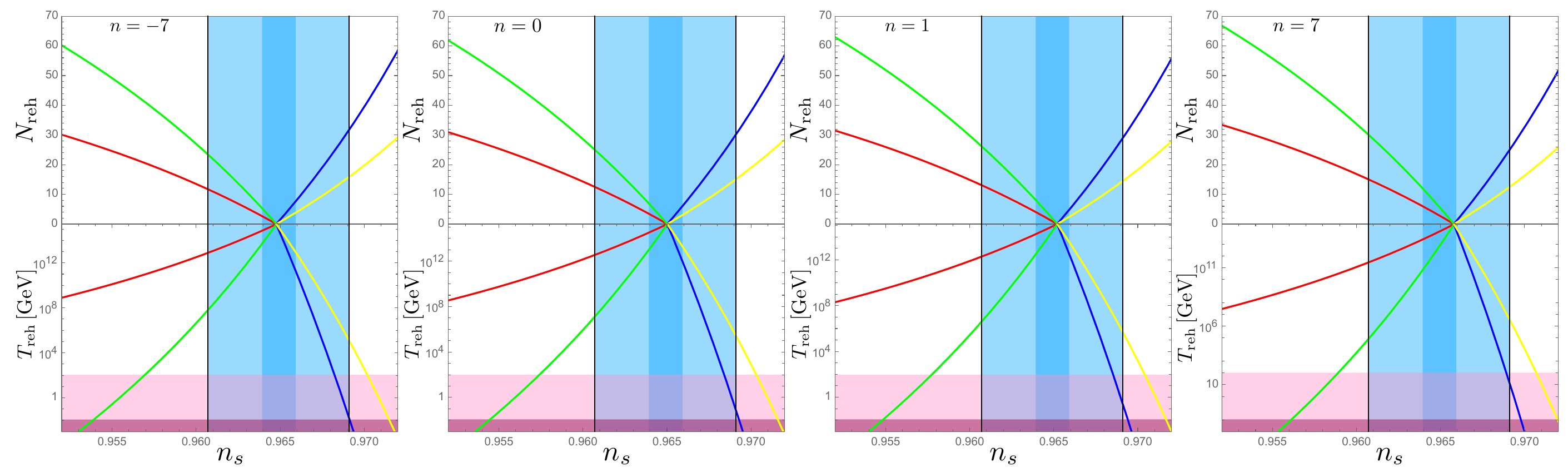}
    \caption{The plots show the extension  of reheating era from $N_\text{reh}$
    and the reheating temperature $T_\text{reh}$ versus  $n_s$ for the model, in  the case in which the parameter $\alpha_1\ll 1$ and $\alpha_2\kappa^{-n/2}\sim\mathcal{O}(1)$. As before in all plots the curves and shaded regions  are as the Fig.(\ref{fig1}).  Here we have considered for $\alpha_1$ and $\alpha_2$ the values $\alpha_1=1\times 10^{-15}$ and $\alpha_2=1\times M^{-n}_\text{pl}$, respectively.}
    
    \label{fig4}
\end{figure}

The curves of the Figs.\ref{fig2}, \ref{fig3} and \ref{fig4} show a similar behavior to the curves of Fig.\ref{fig1}. Here we have considered different values of the parameters $\alpha_1$ and $\alpha_2$ related to the coupling function $f(\phi)$. From these figures we can note that  the results obtained for the duration   of the reheating scenario characterized by number  $N_\text{reh}$ and the reheating temperature in terms of the scalar spectral index do not depend of the values of $\alpha_1$ and 
$\alpha_2$ when these parameters are the order of the unity or much smaller that unity. In this sense, from Planck data at 1$\sigma$ bound on the scalar spectral index, we find that  for values of the EoS parameter $w_\text{reh}\ge 0$ the model presents higher values of temperature $T_\text{reh}$ and the length of  reheating scenario becomes small.

\begin{figure}
    \centering
    \includegraphics[width=18cm]{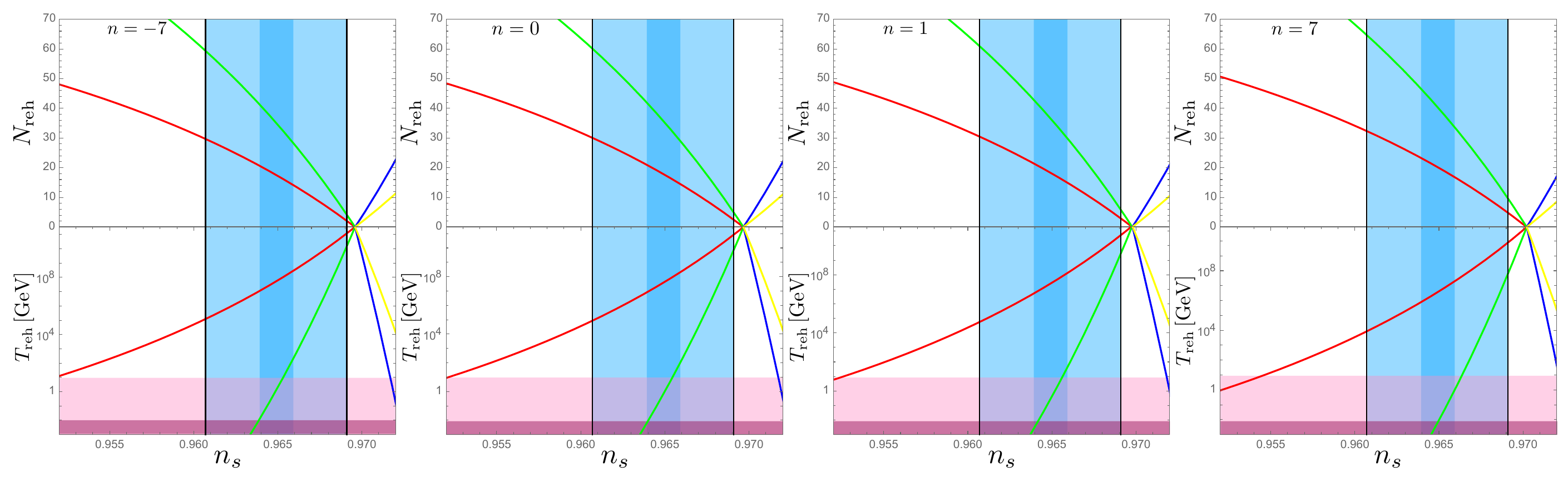}
    \caption{As  before the plots show the number $N_\text{reh}$
    and the temperature $T_\text{reh}$ versus the scalar spectral index $n_s$, for the case in which both parameters $\alpha_1$ and $\alpha_2\kappa^{-n/2}$ are much large than unity, i.e.,  $(\alpha_1,\alpha_2\kappa^{-n/2})\gg 1$. As always in all plots the curves and shaded regions  are as the Fig.(\ref{fig1}). Here we have used for $\alpha_1$ and $\alpha_2$ the values $\alpha_1=10^{15}$ and $\alpha_2=1\times 10^{15}M^{-n}_\text{pl}$, respectively.
    }
    \label{fig5}
\end{figure}

In Fig.\ref{fig5} we show the length of reheating characterized by the number $N_\text{reh}$ and the temperature $T_\text{reh}$ as a function of the scalar spectral index, when  the parameters $\alpha_1$ and $\alpha_2\kappa^{-n/2}$ are much large that the unity i.e., $(\alpha_1,\alpha_2\kappa^{-n/2})\gg 1$. Here we can observe that for the specific values of the EoS parameters $w_\text{reh}=2/3$ and  $w_\text{reh}=1$ ($w_\text{reh}>0$), the observable $n_s$ in both cases is totally excluded by the Planck data at 1$\sigma$ bound on $n_s$. Besides, we note that within the region of Planck data, the EoS parameters  $w_\text{reh}=-1/3$ and $w_\text{reh}= 0$ predict high temperatures during the reheating epoch. Also, we observe that the length of  reheating for the specific case in which the EoS parameter $w_\text{reh}=0$ becomes large in relation to the case when 
$w_\text{reh}=-1/3$. In relation to   the power $n$  associated to the coupling function $f(\phi)$, we note that all curves in the different panels are similar (from $n=-7$ to $n=7$) with which  for values of  $(\alpha_1,\alpha_2\kappa^{-n/2})\gg 1$, the model becomes independent of power $n$ in analogously as it occurs in the another figures. These results suggest that for large values of the parameters $\alpha_1$ and $\alpha_2$ ($\alpha_1\gg 1$ and $\alpha_2\kappa^{-n/2}\gg 1$) and then  large values for the coupling  function   $f(\phi)=\alpha_1+\alpha_2\phi^n\gg1$ (large-$f(\phi)$), the model does not work for all  EoS parameter $w_\text{reh}$ at 1 $\sigma$ bound on $n_s$. Thus, for large-$f(\phi)$, we find that for  positive values of EoS parameter $w_\text{reh}$, the model is excluded from the observational data imposed by  Planck satellite (1$\sigma$ bound on $n_s$).

\section{Concluding remarks}

In this paper we have analyzed the reconstruction in the framework of the modified gravity from an interaction between the scalar field through a coupling function $f(\phi)$ and the trace of the energy-momentum tensor $T$, considering as  cosmological parameter the scalar spectral index $n_s=n_s(N)$, where $N$ corresponds to the number of $e$-folds. By assuming  the slow roll approximation, we have found a general formalism of reconstruction for the background variables in the framework of $f(\phi)T$ inflation. In the context of this general analysis, we have obtained from the attractor $n_s(N)$ an integral  relation for the effective potential in terms of the number of $e$-folds $N$  and also we have  found a relation between the scalar field and the $e$-folds $N$, in order to rebuild  the effective potential as a function of the scalar field.

To reconstruct the effective potential $V(\phi)$, we have assumed a specific form for the coupling function $f(\phi)$ (power-law) together with the 
 simplest example  for the scalar spectral index in terms of the number $N$  given by $n_s(N)=1-2/N$ (standard attractor) for large $N$. From this attractor, we have assumed our general formalism to get the effective potential as a function of the number $N$. 
 From the methodology used for the reconstruction of the scalar potential, we have found that the expression for the effective potential $V(\phi)$ depends of the sign  of the integration constant ($B\lesseqqgtr 0$) obtained  during the first scenario, in which it is necessary to determine the potential as a function of the number of $e$-folds $N$, i.e.,  $V(N)$.  Also, we have obtained  that the reconstruction of the effective potential $V(\phi)$ depends on  the coupling function $f(\phi)$ associated to the interaction between this function and  the trace of energy momentum tensor $T$ from the term $f(\phi)\,T$ in the action given by Eq.(\ref{accion}). 
 In  particular for the specific case in which the parameters $\alpha_1=\alpha_2=0$ (or $f(\phi)=0$) the reconstructed  effective potential is reduced to the found   in the framework of the GR. In Fig. \ref{fig1a} we show the evolution of the reconstructed  dimensionless effective potential $\bar{V}(\phi)=B\,V(\phi)$ as a function of the scalar field from Eq.(\ref{Vff}), in the specific case in which the integration constant $B$ is positive. In this plot, we have considered different values of the power $n$ related to the coupling function $f(\phi)$. 
 
In relation to study of the reheating epoch for our model in the framework of $f(\phi)T$ inflation, we have found that it is possible parameterize this era in terms of the reheating parameters such as;  number of $e-$folds and the temperature during the reheating together with the EoS parameter $w_\text{reh}$. 
In this context,  we have been able to determine the reheating parameters; temperature and the number of $e-$folds during the reheating era as a function of parameters present in the model, the EoS parameter $w_\text{rh}$ and  
 the observational parameters such as; the scalar power spectrum, scalar index and the tensor to scalar ratio, see Eqs.(\ref{I1}) and (\ref{I2}).
It is important to notice that from Figs.\ref{fig1}, \ref{fig2}, \ref{fig3} and \ref{fig4} in the which we have considered the specific cases in which the parameters $\alpha_1$ and $\alpha_2$ are;  ($\alpha_1,\alpha_2\kappa^{-n/2}$)$\ll$ 1 or ($\alpha_1,\alpha_2\kappa^{-n/2})\sim\mathcal{O}(1)$, the temperature and the number of $e$-folds during the reheating era do not depend of the power $n$ related to the coupling function $f(\phi)$, since these figures show similar results. Here we have used different values from $n=-7$ to $n=7$. Thus, we have observed that 
 any temperature between the BBN limit and the instantaneous reheating value is  permitted inside the Planck's 1$\sigma$ limit regardless of the values of EoS parameter $w_\text{reh}$ and the power $n$, excepting the negative value  $w_\text{reh}=-1/3$. From the Planck data (1$\sigma$ bound on $n_s$) we have found that the model for the cases in which the EoS parameter  $w_\text{reh}\ge 0$ predicts higher values associated to the reheating temperatures, in particular for $w_\text{reh}=0$ we have  $T_\text{reh}\gtrsim 5\times10^{7}$GeV. 
Besides, we have obtained that the duration of the reheating stage characterized by   the number of $e$-folds during this epoch becomes  small in which the number of $e-$ folds  $N_\text{reh}\lesssim 30$ for the different reheating temperatures from the constraints imposed by Planck data (1$\sigma$ bound on $n_s$).

On the other hand, in the specific case in which the parameters $\alpha_1$ and $\alpha_2$ related to the coupling function $f(\phi)$ are $\alpha_1\gg 1$ and $\alpha_2\kappa^{-n/2}\gg 1$ i.e., large-$f(\phi)$ (see Fig.\ref{fig5}), we have found that for positive values of the EoS parameter related to reheating i.e.,  $w_\text{reh}>0$, the curves associated to the temperatures and numbers  of $e-$folds during the reheating stage are excluded by Planck data at 1$\sigma$ bound on the index $n_s$. From Fig.\ref{fig5} we have noted that 
 results for the quantities $T_\text{reh}$ and $N_\text{reh}$ suggest that for large values of the parameters $\alpha_1$ and $\alpha_2$  and then  large values for the coupling  function   $f(\phi)=\alpha_1+\alpha_2\phi^n\gg1$, the model does not work for all  EoS parameter $w_\text{reh}$ at 1 $\sigma$ bound on $n_s$. Thus, for large-$f(\phi)$, we have found  that for  positive values of EoS parameter $w_\text{reh}$, the model is excluded from the observational data imposed by  Planck satellite (1$\sigma$ bound on $n_s$). 

Finally in this article, we have not addressed the 
reheating analysis (numerical) considering  an EoS parameter 
that depends on time i.e., $w_\text{reh}(t)$, in order to determine the temperature and duration of the reheating era. In this
context, we hope to return to this point in the near future.

\section{acknowledgments}
C. R. thanks to the Vicerrector\'ia de Investigaci\'on y Estudios Avanzados, Pontificia Universidad Cat\'olica de Valpara\'iso, for  Postgraduate Scholarship PUCV 2019.
\\
\\
\\
\\
\\

\end{document}